\documentclass[12pt]{iopart}

\usepackage{graphicx}
\usepackage{dcolumn}
\usepackage{bm}
\usepackage{hyperref}
\hypersetup{colorlinks=true, citecolor=blue, urlcolor=blue, linkcolor=blue}
\usepackage{color}
\usepackage{dsfont}
\usepackage{subfigure}
\usepackage{graphicx}
\usepackage{epsfig}
\usepackage{epstopdf}
\usepackage[english]{babel}
\usepackage[T1]{fontenc}
\usepackage[utf8]{inputenc}
\usepackage[normalem]{ulem}

\newcommand {\beq} {\begin{eqnarray}}
\newcommand {\eeq} {\end{eqnarray}}
\newcommand {\eeqn} [1] {\label{#1} \end{eqnarray}}

\begin{document}

\title[]{Tracing the dynamical interplay of low-energy reaction processes of exotic nuclei using a two-center molecular continuum}

\author{Laura Moschini and Alexis Diaz-Torres}

\address{Department of Physics, Faculty of Engineering and Physical Sciences, University of Surrey, Guildford, Surrey GU2 7XH, United Kingdom}

\ead{l.moschini@surrey.ac.uk, a.diaztorres@surrey.ac.uk}

\begin{abstract}
The competition among reaction processes of a weakly-bound projectile at intermediate times of a slow collision has been unraveled. This has been done using a two-center molecular continuum within a semiclassical, time-dependent coupled-channel reaction model. Dynamical probabilities of elastic scattering, transfer and breakup agree with those derived from the direct integration of the time-dependent Schr\"odinger equation, demonstrating the usefulness of a two-center molecular continuum for gaining insights into the reaction dynamics of exotic nuclei.  
\end{abstract}


\textit{Introduction.} Understanding the physics of low-energy nuclear reactions of exotic nuclei is crucial for the experimental programmes using re-accelerated rare-isotope beams at new nuclear research facilities such as FRIB (USA), RIKEN-RIBF (Japan) and FAIR (Germany). Exotic nuclei are mostly weakly bound and their low-energy direct reactions involve the interplay between different channels: elastic and inelastic scattering, transfer, and breakup. An effective theoretical description of such processes has to take into account continuum states \cite{AUSTERN1987125}. Time-dependent approaches are often used to disentangle the reaction mechanism \cite{MILEK198565,BDTPRC92}, so in Ref.\ \cite{MMVPRC103} a simple model, that assumes semiclassical relative motion and neglects angular coordinates, was used to understand how the continuum impacts on direct reactions of one-neutron halo nuclei. In particular, a coupled-channels solution involving different discretised continuum configurations was compared to the numerical solution of the time-dependent Schr\"odinger equation (TDSE). The use of sets of continuum states which are sensitive to the phase shift induced by only one nucleus was found to be inaccurate, especially in the case of a dominant breakup channel. Also, due to the technical design for the continuum inclusion (i.e., non-unitarity of the time-evolution operator) the model can only calculate the asymptotic result, thus preventing the possibility to follow the reaction during its evolution. As a possible solution, the use of a two-center model was suggested. This would involve the definition of a set of discretised ``molecular'' pseudostates that have a dynamical phase shift caused by two potential wells, each of which is associated with either the projectile or the target. The two-center description of reactions is a well established technique in atomic physics \cite{DELOS81,FRITSCHLIN19911}, that has been applied to describe molecular single-particle (bound) states in nuclear processes \cite{TSFG78,MRPLB157,NSPPRC35, KGS91,DTTS2002,DIAZTORRES2005373,ADTprl101}. However, the inclusion of a two-center molecular single-particle continuum in reaction theory of weakly-bound nuclei is a novel feature. The central objective of the present paper is to introduce a discretised molecular continuum and to study its role in a time-dependent semiclassical framework \cite{Fallot,Marta}, similar to that used in Ref.~\cite{MMVPRC103}, allowing us to trace the dynamical interplay of all reaction channels in direct reactions. 
We consider a valence neutron that is initially in a loosely bound state of a projectile $P$ impinging on a target $T$. In this model, the $P$-$T$ motion is treated by classical mechanics, whereas the motion of the valence neutron relative to the overall center-of-mass of the $P$-$T$ system is described by quantum mechanics. This system undergoes a direct reaction, so the valence neutron is finally expected to be either in a bound state of $P$ or $T$ (elastic and transfer channels, respectively), or in the continuum of scattering states (breakup channels). \\

\begin{figure}[!h]
    \centering
    \includegraphics[width=\columnwidth]{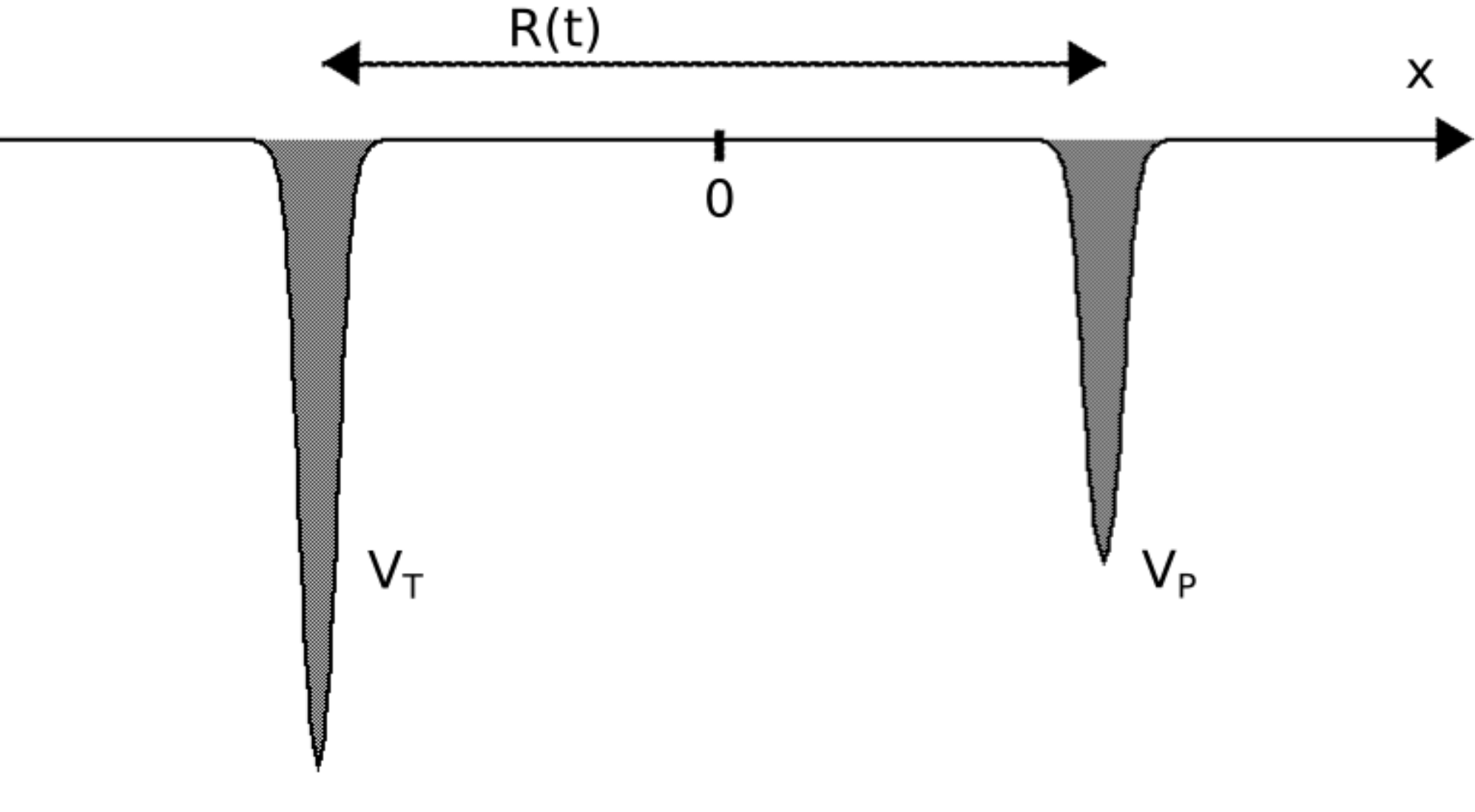}
   \caption{Schematic representation of projectile ($V_P$) and target ($V_T$) potential wells on the one-dimensional spatial grid, labelled by the variable $x$. The origin of the x-axis is located in the overall center-of-mass point. The parameter $R(t)$ measures the distance between the minima of the two potentials.}
    \label{Fig0}
\end{figure}

\textit{Methods.} To follow the time evolution of a valence neutron in direct reactions, the TDSE equation has to be solved:
\begin{equation}
    i \hbar \frac{\partial}{\partial t}\Psi(x,t) = {\hat H} \Psi(x,t).
    \label{TDSE}
\end{equation}
The wavefunction $\Psi(x,t)$ describes the probability amplitude to find the valence neutron at time $t$ on the one-dimensional spatial grid, being $x$ the position variable of the neutron relative to the origin of the coordinate system that is located at the overall centre-of-mass point.  The scattering of the valence neutron in the field of two potential wells representing target or projectile nuclei of mass numbers $A_i$, with $i=T,P$, is given by the single-particle Hamiltonian:
\begin{equation}
    {\hat H}(x,R(t)) = -\frac{\hbar^2}{2 m_0}\frac{d^2}{dx^2} + V_T\left(x-\frac{\mu}{A_T}R(t)\right)+V_P\left(x+\frac{\mu}{A_P}R(t)\right).
    \label{HTC}
\end{equation}
Here, $m_0$ is the neutron mass, $\mu=A_TA_P/(A_T+A_P)$ is the the dimensionless $P$-$T$ reduced mass, and the potentials $V_i$ are expressed in the overall center-of-mass frame as a function of the internuclear distance $R(t)$, a parameter that measures the distance between the minima of the two potential wells on the one-dimensional axis.
The two wells are assumed to have a (static) Woods-Saxon shape, and their relative distance follows a fixed trajectory, $R(t) = R_{min} + \dot{R}\,t$, 
with a distance of minimal approach $R_{min}$ and a constant radial velocity $\dot{R} = \pm \sqrt{2 E/\mu}$, $E$ being the incident energy. Both the radial velocity and time  are considered negative (positive) for the ingoing (outgoing) branches of the trajectory. A schematic representation of the spatial grid with the two potential wells at a distance $R(t)$ is displayed in Fig.~\ref{Fig0}.

Using the wave-function expansion method, the time-dependent wave function in Eq.~(\ref{TDSE}) is represented by a truncated basis set of two-center molecular states, $\{\Phi_n\}$, with eigenenergies $E_n(R(t))$:
\begin{equation}
    \Psi(x,t) = \sum_{n=1}^{N} c_n(t) \Phi_n(x,R(t))e^{-i E_n(R(t)) t /\hbar},
    \label{TDexp}
\end{equation}
and the time-dependent coupled-channel equations are solved for the amplitudes, $c_n(t)$:
\begin{equation}
           \dot{c}_n(t) = - \dot{R}(t) \sum_m c_m(t) \, T_{nm}(R(t)) \, e^{i(E_n(R(t))-E_m(R(t)))t/\hbar},
           \label{CC}
\end{equation}
where the dot indicates a time derivative and the matrix elements $T_{nm}(R(t))=\langle \Phi_n(x,R(t)) | \frac{\partial}{\partial R} | \Phi_m(x,R(t))  \rangle$ are the radial derivative couplings between the molecular basis states. Eq.~(\ref{CC}) is the fundamental equation for the two-center coupled-equations (TCCE) method. 
The two-center basis states, $\Phi_n(x,R(t))$, are obtained by diagonalizing the system Hamiltonian (\ref{HTC}) for each $R(t)$
in a truncated box basis set  $\{\varphi_{\nu}\}$ \cite{LandauQM}, the fixed grid length $L$ being the size of the box. Therefore, at each internuclear distance, $R(t)$, we obtain a set of eigenvectors 
$ \Phi_n(x,R(t)) = \sum_{\nu=1}^{M} a_{n,\nu}^{R(t)} \varphi_{\nu}(x)$. This set is composed of bound states, with each state containing a negative energy,
and a finite number of continuum pseudostates with a positive energy.  Pseudostates, like actual continuum states, present an oscillatory behaviour and a phase shift caused by the two potential wells; unlike true continuum they are discretised and normalisable (i.e.,\ their wave function is zero at the box edges). 
The set of coefficients, $\{a_{n,\nu}^{R(t)}\}$, has an arbitrary phase, $e^{\pm i \pi}$, so at each time step, we apply the method of least squares and minimize the quantity $ S_n = \sum_{\nu=1}^M \left(a_{n,\nu}^{R(t-dt)}-a_{n,\nu}^{R(t)}\right)^2$ for each $n$-th state. 
This reduces the arbitrary variation of the sign of $\{a_{n,\nu}^{R(t)}\}$ with small variations of $R(t)$ and thus makes the matrix elements of the radial derivative coupling, $T_{nm}(R(t))$,
 a monotonically changing function as the internuclear radius varies, which is physically expected \cite{NSPPRC35}.

The two-center basis is orthonormal for each $R(t)$, i.e.,  $\langle \Phi_n(x,R(t))| \Phi_m (x,R(t)) \rangle = \delta_{nm}$. So, the matrix element $T_{nm}(R(t))$ takes the form: 
\begin{equation}
    T_{nm}(R(t)) = \sum_{\nu=1}^{M} \frac{a_{m,\nu}^{R(t-dt)}-a_{m,\nu}^{R(t+dt)}}{ R(t-dt)-R(t+dt)} \langle \Phi_n(x,R(t)) |   \varphi_{\nu}(x) \rangle,
    \label{Tnm}
\end{equation}
where $dt$ is the time step for the time evolution.

To solve Eq.~(\ref{CC}), we use the Chebyshev propagator \cite{KARPC45}
which consists of representing the time evolution operator as a convergent series of Chebyshev polynomials (see  Appendix C of Ref.\ \cite{DTWPRC97}).
As the initial state, we selected one of the bound states of a separated potential well, the projectile's ground state in our test case. At the beginning of the collision, for a large internuclear distance, it coincides with one molecular bound state. Hence the initial condition for Eq.~(\ref{CC}) is $c_n(t_i)=\delta_{nk}$, $k$ being the label denoting the ground state of the projectile and $t_i$ referring to the initial time. The initially unoccupied bound state of the target nucleus, which will be associated with the neutron transfer process, is labelled by $j$.  The probability to find the neutron in the nth state at time $t$ is:
\begin{equation}
{\cal P}^{TCCE}_n(t) = |c_n(t)|^2.
\label{TCCEprob}
\end{equation}

To check the reliability of this approximate coupled-channels solution, we have compared it with the direct numerical integration of the Schr\"odinger equation (\ref{TDSE}), carried out with the Runge-Kutta fourth-order method \footnote{We have also tested both the Runge-Kutta and the Chebyshev propagator against the Pad\'e approximation of the time-evolution operator \cite{MMVPRC103}, an alternative yet most computationally demanding algorithm, obtaining an excellent agreement in each case.}. In this case the probability to find the neutron in a particular channel at time $t$ is:
\begin{equation}
    {\cal P}^{TDSE}_n(t) = |\langle \Psi(x,t) | \Phi_n(x,R(t)) e^{-i E_n(R(t)) t /\hbar} \rangle|^2.
    \label{TDSEprob}
\end{equation}

\textit{Numerical details.} Due to the unitarity of the time evolution operator, our model provides stable solutions: the norm of $\Psi(x,t)$ and $\sum_n |c_n(t)|^2$ are preserved with an accuracy of $\sim 10^{-5}$ during the entire collision process. This is particularly interesting because it solves an important issue pointed out in Ref. \cite{MMVPRC103}: following the probability for the valence neutron to occupy each molecular state (\ref{TCCEprob}), we are now able to unravel the dynamical interplay of reaction channels during the entire collision process.
\begin{table}[h!]
\centering
\caption{\label{tab} Mass numbers A, depth $V_0$, range $r_0$, and diffuseness $a_0$ of the Woods-Saxon potentials related to the projectile and target nuclei.}
 \begin{tabular}{l c c c c} 
 &&&&\\
 \hline
 Nuclei & $A$ & $V_0$ (MeV) & $r_0=1.2\cdot A^{1/3}$ (fm) & $a_0$ (fm) \\ 
 \hline
 Projectile & 6 & -2.0 & 2.2 & 0.7\\
 Target     & 5 & -3.0 & 2.1 & 0.7\\
 \hline
 \end{tabular}
\end{table}

As a qualitative test case we have chosen initial conditions that mimic the collision of a weakly-bound one-neutron halo nucleus with a target. In Table~\ref{tab} we list the mass numbers and Woods-Saxon potential parameters associated with the projectile and target nuclei. These values do not correspond to real colliding nuclei because the present model is only qualitative and does not describe realistic collisions. The potential well of the projectile, $V_P$, has one weakly-bound state of energy $E_P=-0.54$~MeV, initially occupied by the valence neutron, while the potential well of the target, $V_T$, has one unoccupied bound state of energy $E_{T}=-0.96$~MeV. To build the molecular basis we have chosen $M=500$ box states defined on a $L=\pm 350$~fm grid with $dx=0.3$~fm, and we have included the first $N=50$ molecular states in the calculation. We set an initial internuclear distance of $80$~fm, and a distance of minimal approach $R_{min}=8$~fm.\\

\begin{figure}[!h]
    \centering
    \includegraphics[width=\columnwidth]{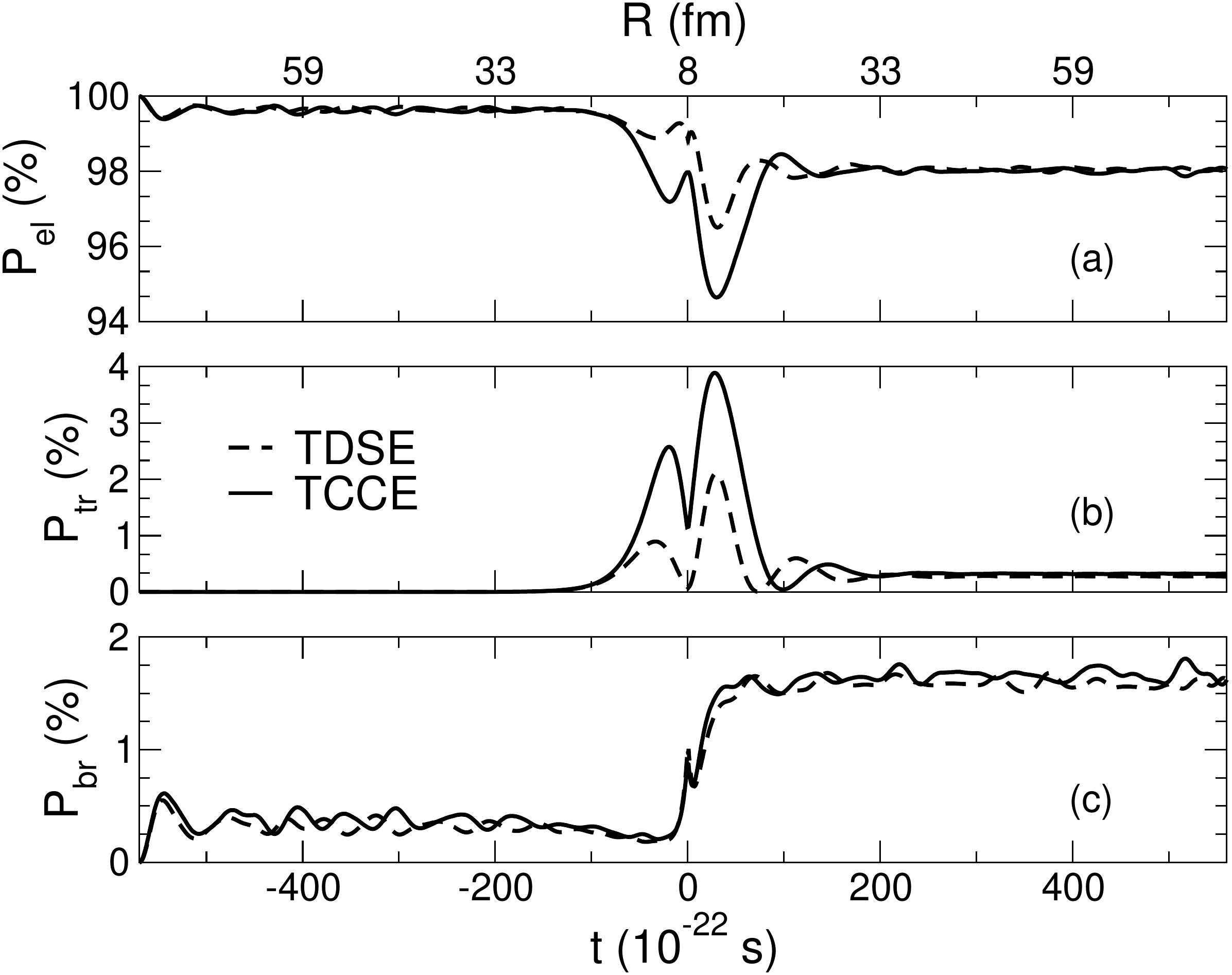}
   \caption{Probability for (a) elastic, (b) transfer, and (c) breakup channels as a function of time and internuclear distance $R$, using the two-center coupled-equations method (solid line) compared to the direct integration of the time-dependent Schr\"odinger equation (dashed line) for a collision at $E=0.05$~MeV. The distance of minimal approach ($R_{min} = 8$~fm) occurs at $t=0 \times 10^{-22}$ s.}
    \label{Fig1}
\end{figure}

 \textit{Results and discussion.} In Fig.~\ref{Fig1} the TCCE probability for elastic scattering ${\cal P}_{el}(t)=|c_k(t)|^2$, transfer ${\cal P}_{tr}(t)=|c_j(t)|^2$, and breakup channels ${\cal P}_{br}(t)=\sum_{n\neq j,k}^N |c_n(t)|^2$ as a function of time is compared to the TDSE results, using solid and dashed lines respectively. In this case we consider an initial energy of $E=0.05$~MeV and a time step $dt=0.5\cdot 10^{-22}$~s. The excellent comparison with TDSE calculation confirms the accuracy of our TCCE model. The small oscillation of the breakup probability as a function of time is due to the effect of the moving basis, which depends on the collision energy. In the semiclassical coupled-channel theory based on molecular states, the moving basis causes non-zero, asymptotic radial derivative couplings. This is well recognised in molecular theory of atomic collisions, where phenomenological electron translational factors are introduced to remove such unphysical couplings \cite{FRITSCHLIN19911}. 
 In presence of oscillations, the final probability for each channel can be determined by the average value over a period of oscillation of the asymptotic results. The same quantity can be provided by several traditional reaction methods based on a one-center continuum, e.g.\ the continuum-discretised coupled-channels calculations (CDCC) \cite{AUSTERN1987125}. However, they are unable to describe what happens at intermediate times of a collision, because they are either time-independent models or do not preserve the total probability due to non-unitarity or over-completeness of the basis \cite{MMVPRC103}. Instead, a molecular time-dependent basis is able to simultaneously show the probability of each reaction channel at each time. 
 
\begin{figure}[!h]
    \centering
    \includegraphics[width=\columnwidth]{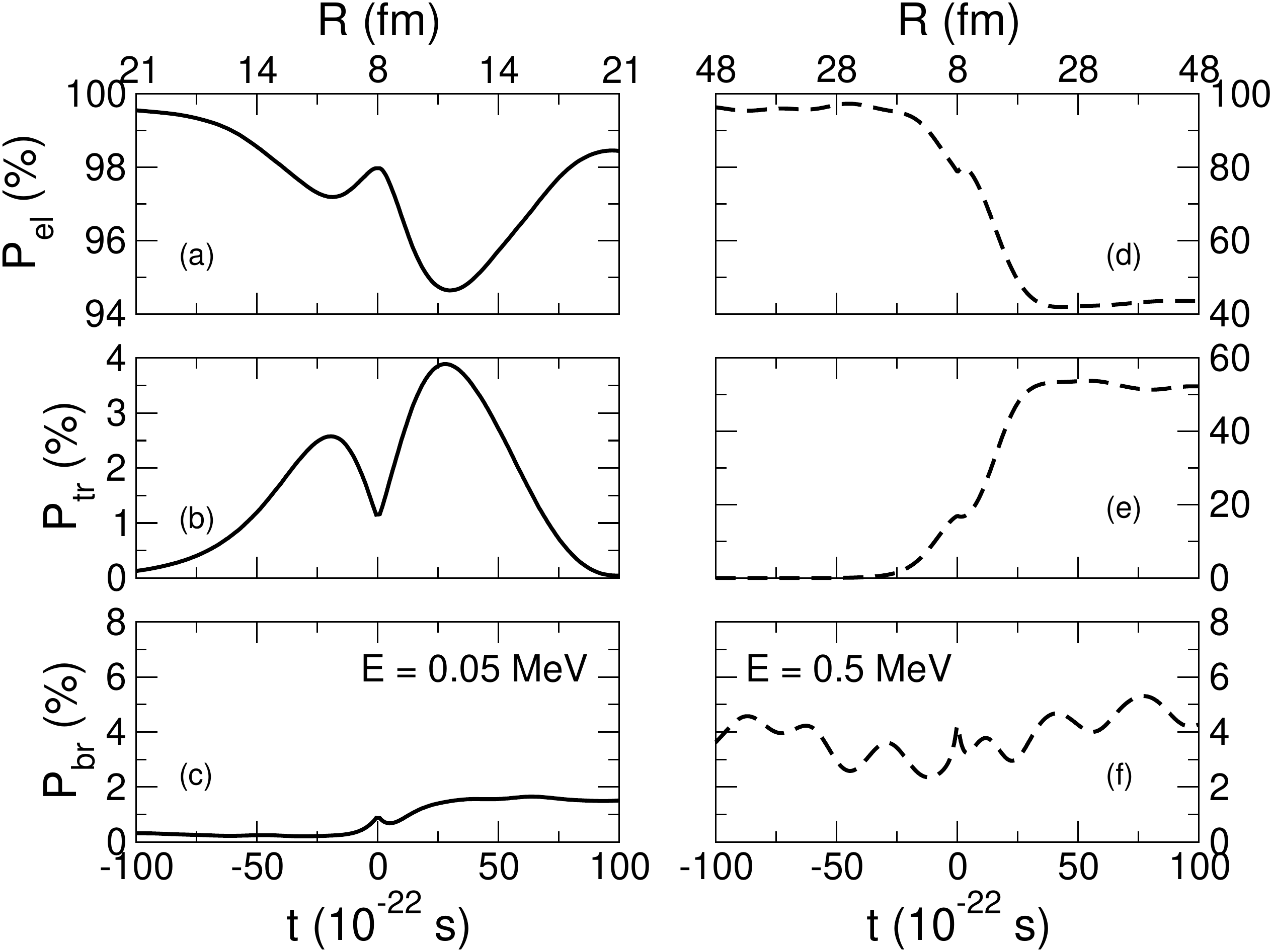}
   \caption{Probability for (a,d) elastic, (b,e) transfer, and (c,f) breakup channels as a function of time and internuclear distance $R$ for two collision energies: $E=0.05$~MeV (solid line,  panels (a)-(c)) and $E=0.5$~MeV (dashed line, panels (d)-(f)), using the two-center coupled-equations method. Compared to Fig.~\ref{Fig1}, the time window is focused on internuclear distances around the distance of minimal approach ($R_{min} = 8$~fm). The elastic and transfer probabilities are presented in different scales.}
    \label{Fig2}
\end{figure}

 Fig.~\ref{Fig2} shows the TCCE probability for each reaction channel of the collision at $E=0.05$~MeV (solid line) and $E=0.5$~MeV (dashed line), for which we used time steps of $0.5\cdot 10^{-22}$~s and $0.16\cdot 10^{-22}$~s, respectively. 
 These two different time steps provide the same $|R(t-dt)-R(t+dt)|=|2dR|=0.13$~fm value, which is crucial for an accurate calculation of the $T_{nm}$ matrix elements in Eq.~(\ref{Tnm}). In Fig.~\ref{Fig2}, we highlight a narrow time window around the turning point, using different scales for elastic and transfer probabilities.
 We observe how the reaction at lower energy is mainly an elastic scattering, while with increasing energy, transfer and breakup processes become more important. This can be explained by the fact that the larger the incident energy, the larger the radial velocity $\dot{R}$ in Eq.\ (\ref{CC}), and the effect that large range  $T_{nm}$ matrix elements (associated with the breakup channels) have on excitations becomes significant. We can also discover how, in the collision at lower energy, there is enough time for fast rearrangements and the valence neutron is shared by the molecular bound states (a transitional nuclear molecule \cite{NMbook}), subsequently the probability to find the neutron in the continuum increases.
 Based on this result we can also solve another issue discussed in Ref.\ \cite{MMVPRC103}, and we conclude that molecular continuum is the appropriate tool to describe processes with higher breakup components. This is due to technical properties of the two-center molecular basis: (i) it is the only complete set of states included in the calculation, thus avoiding over-completeness issues, (ii) its Hilbert space spans the entire spatial region where the collision takes place, not only narrow spatial ranges around the individual potentials, so any component of the total neutron wave function can be described by the molecular states, and  (iii) the molecular basis includes the phase shifts induced by both potentials, providing a more realistic description of the particle in the field of both potential wells simultaneously, as opposed to alternative bases restricted to the separated individual potentials.
\\ 

\textit{Summary.} We have introduced the notion of a two-center molecular continuum and have demonstrated its usefulness within a simple dynamical reaction model for a weakly-bound projectile. The use of a two-center molecular continuum solves the problem of physical interpretation of reaction processes at intermediate times, allowing one to unravel the dynamical competition among direct reaction processes (elastic and inelastic scattering, transfer, and breakup) during the entire collision process. This is an improvement on solely asymptotic calculations, such as simulations based on a one-center continuum \cite{MMVPRC103}. The good outcomes of the present calculations push our investigation towards the refinement of the two-center molecular continuum basis along with the development of a realistic, three-dimensional quantitative reaction model.

\ack

This work has received funding from the United Kingdom Science and Technology Facilities Council (STFC) under Grants No.\  ST/L005743/1 and ST/P005314/1.

\section*{References}
%

\end{document}